\documentclass[twocolumn,showpacs,preprintnumbers,superscriptaddress,amsmath,amssymb]{revtex4}

\usepackage{graphicx} 
\usepackage{dcolumn}
\usepackage{bm}
\usepackage{epsfig}
\usepackage{figlatex}
\usepackage{pdflscape}
\usepackage{subfigure} 
\usepackage{multirow}
\usepackage[colorlinks=true,citecolor=blue,filecolor=blue,linkcolor=blue,urlcolor=blue,pdftex]{hyperref}

\begin{document}

\title{The neutron background of the XENON100 dark matter search experiment}

\newcommand{\versionauthors}{2013-01-07}
\newcommand{\versionmembers}{2012-11-22}
\newcommand{\papershorttitle}{Neutron Background Paper}


\newcommand{\assergi}{\affiliation{INFN, Laboratori Nazionali del Gran Sasso, Assergi, 67100, Italy}}
\newcommand{\bern}{\affiliation{Albert Einstein Center for Fundamental Physics, University of Bern, Sidlerstrasse 5, 3012 Bern, Switzerland}}
\newcommand{\bologna}{\affiliation{University of Bologna and INFN-Bologna, Bologna, Italy}}
\newcommand{\columbia}{\affiliation{Physics Department, Columbia University, New York, NY 10027, USA}}
\newcommand{\coimbra}{\affiliation{Department of Physics, University of Coimbra, R. Larga, 3004-516, Coimbra, Portugal}}
\newcommand{\heidelberg}{\affiliation{Max-Planck-Institut f\"ur Kernphysik, Saupfercheckweg 1, 69117 Heidelberg, Germany}}
\newcommand{\houston}{\affiliation{Department of Physics and Astronomy, Rice University, Houston, TX 77005 - 1892, USA}}
\newcommand{\laquila}{\affiliation{Department of Physics, University of L'Aquila, 67010, Italy}}
\newcommand{\losangeles}{\affiliation{Physics \& Astronomy Department, University of California, Los Angeles, USA}}
\newcommand{\mainz}{\affiliation{Institut f\"ur Physik, Johannes Gutenberg-Universit\"at Mainz, 55099 Mainz, Germany}}
\newcommand{\munster}{\affiliation{Institut f\"ur Kernphysik, Wilhelms-Universit\"at M\"unster, 48149 M\"unster, Germany}}
\newcommand{\nikhef}{\affiliation{Nikhef  and the University of Amsterdam, Science Park, Amsterdam, Netherlands}}
\newcommand{\purdue}{\affiliation{Department of Physics, Purdue University, West Lafayette, IN 47907, USA}}
\newcommand{\shanghai}{\affiliation{Department of Physics, Shanghai Jiao Tong University, Shanghai, 200240, China}}
\newcommand{\subatech}{\affiliation{SUBATECH, Ecole des Mines de Nantes, CNRS/In2p3, Universit\'e de Nantes, 44307 Nantes, France}}
\newcommand{\torino}{\affiliation{INFN-Torino and Osservatorio Astrofisico di Torino, 10100 Torino, Italy}}
\newcommand{\weizmann}{\affiliation{Department of Particle Physics and Astrophysics, Weizmann Institute of Science, 76100 Rehovot, Israel}}
\newcommand{\zurich}{\affiliation{Physics Institute, University of Z\"{u}rich, Winterthurerstr. 190, CH-8057, Z\"{u}rich, Switzerland}}

\author{E.~Aprile}\columbia 
\author{M.~Alfonsi}\nikhef
\author{K.~Arisaka}\losangeles
\author{F.~Arneodo}\assergi
\author{C.~Balan}\coimbra
\author{L.~Baudis}\zurich
\author{B.~Bauermeister}\mainz
\author{A.~Behrens}\zurich
\author{P.~Beltrame}\weizmann\losangeles
\author{K.~Bokeloh}\munster
\author{A.~Brown}\purdue
\author{E.~Brown}\munster
\author{G.~Bruno}\assergi
\author{R.~Budnik}\columbia 
\author{J.~M.~R.~Cardoso}\coimbra
\author{W.-T.~Chen}\subatech
\author{B.~Choi}\columbia
\author{A.~P.~Colijn}\nikhef
\author{H.~Contreras}\columbia
\author{J.~P.~Cussonneau}\subatech
\author{M.~P.~Decowski}\nikhef
\author{E.~Duchovni}\weizmann
\author{S.~Fattori}\mainz
\author{A.~D.~Ferella}\assergi\zurich
\author{W.~Fulgione}\torino
\author{F.~Gao}\shanghai
\author{M.~Garbini}\bologna
\author{C.~Ghag}\losangeles
\author{K.-L.~Giboni}\columbia
\author{L.~W.~Goetzke}\columbia
\author{C.~Grignon}\mainz
\author{E.~Gross}\weizmann
\author{W.~Hampel}\heidelberg
\author{F.~Kaether}\heidelberg
\author{A.~Kish}\email[]{Corresponding author: alexkish@physik.uzh.ch}\zurich
\author{J.~Lamblin}\subatech
\author{H.~Landsman}\weizmann
\author{R.~F.~Lang}\purdue
\author{M.~Le~Calloch}\subatech
\author{C.~Levy}\munster
\author{K.~E.~Lim}\columbia
\author{Q.~Lin}\shanghai
\author{S.~Lindemann}\heidelberg
\author{M.~Lindner}\heidelberg
\author{J.~A.~M.~Lopes}\coimbra
\author{K.~Lung}\losangeles
\author{T.~Marrod\'an~Undagoitia}\heidelberg\zurich
\author{F.~V.~Massoli}\bologna
\author{A.~J.~Melgarejo~Fernandez}\columbia
\author{Y.~Meng}\losangeles
\author{M.~Messina}\columbia
\author{A.~Molinario}\torino
\author{K.~Ni}\shanghai
\author{U.~Oberlack}\mainz
\author{S.~E.~A.~Orrigo}\email[]{Present address: IFIC, CSIC-Universidad de Valencia, E-46071 Valencia, Spain}\coimbra
\author{E.~Pantic}\losangeles
\author{R.~Persiani}\bologna
\author{G.~Plante}\columbia
\author{N.~Priel}\weizmann
\author{A.~Rizzo}\columbia
\author{S.~Rosendahl}\munster
\author{J.~M.~F.~dos Santos}\coimbra
\author{G.~Sartorelli}\bologna
\author{J.~Schreiner}\heidelberg
\author{M.~Schumann}\bern\zurich
\author{L.~Scotto~Lavina}\subatech
\author{P.~R.~Scovell}\losangeles
\author{M.~Selvi}\bologna
\author{P.~Shagin}\houston
\author{H.~Simgen}\heidelberg
\author{A.~Teymourian}\losangeles
\author{D.~Thers}\subatech
\author{E.~Tziaferi}\zurich
\author{O.~Vitells}\weizmann
\author{H.~Wang}\losangeles
\author{M.~Weber}\heidelberg
\author{C.~Weinheimer}\munster

\collaboration{The XENON100 Collaboration}\noaffiliation

\begin{abstract}
The XENON100 experiment, installed underground at the Laboratori Nazionali del Gran Sasso (LNGS), aims to directly detect dark matter in the form of Weakly Interacting Massive Particles (WIMPs) via their elastic scattering off xenon nuclei. This paper presents a study on the nuclear recoil background of the experiment, taking into account neutron backgrounds from ($\alpha$,n) and spontaneous fission reactions due to natural radioactivity in the detector and shield materials, as well as muon-induced neutrons. 
Based on Monte Carlo simulations and using measured radioactive contaminations of all detector components, we predict the nuclear recoil backgrounds for the WIMP search results published by the XENON100 experiment in 2011 and 2012, 0.11$^{+0.08}_{-0.04}$~events and 0.17$^{+0.12}_{-0.07}$~events, respectively, and conclude that they do not limit the sensitivity of the experiment.
\end{abstract}

\keywords{Dark Matter, Xenon, Background, Neutrons, Monte Carlo Simulation} 
\pacs{95.35.+d, 29.40.-n, 34.80.Dp}

\maketitle

\section{Introduction}
\label{intro}

The XENON100 detector aims at the direct detection of dark matter in the form of Weakly Interacting Massive Particles (WIMPs), and is taking data at the Laboratori Nazionali del Gran Sasso (LNGS) in Italy. It is a double phase (liquid-gas) time projection chamber (TPC) with 62~kg of liquid xenon (LXe) in the active volume viewed by two photomultiplier tube (PMT) arrays on the top and bottom. 
The design of the detector and its working principle are described in detail in Ref.~\cite{xe100-instrument}, and the data analysis procedure is explained in Ref.~\cite{xe100-analysis}. To-date XENON100 is the most sensitive detector for direct dark matter detection, and has set the most stringent limits on the spin-independent elastic WIMP-nucleon scattering for WIMP masses above 8~GeV/c$^{2}$, with a minimum cross section of 2$\times$10$^{-45}$~cm$^{2}$ at 55~GeV/c$^{2}$ (at 90$\%$ confidence level)~\cite{xe100-run10}, and on the spin-dependent scattering for WIMP masses above 6~GeV/c$^{2}$, with a minimum cross section of 3.5$\times$10$^{-40}$~cm$^{2}$ at a WIMP mass of 45~GeV/c$^{2}$, at 90$\%$ confidence level~\cite{xe100-spindependent}.

A WIMP is expected to elastically scatter off a nucleus in the target, producing a low energy nuclear recoil (NR)~\cite{wimps}. 
There are two types of background for a dark matter search with the xenon-based detectors: NRs from hadronic interactions of neutrons, and electronic recoils (ERs) from electromagnetic interactions of $\gamma$-rays and electrons. The different ionization density characteristic of a NR and an ER results in a different probability of electron-ion pair recombination, and thus a different ratio of scintillation to ionization~\cite{AprileDoke}. This provides the possibility of rejecting ER background, which is performed in XENON100 with an efficiency higher than 99\% at $\sim$50\% NR acceptance~\cite{xe100-independent1, xe100-independent2, xe100-run10}.
 
Neutrons can produce single NRs via elastic scattering off xenon nuclei and generate a signal which is, on an event-by-event basis, indistinguishable from that of WIMPs. In addition, fast neutrons are more penetrating than $\gamma$-rays in LXe. It is therefore crucial to minimize and accurately characterize this potentially dangerous background. 
Due to intrinsic contamination with $^{238}$U, $^{235}$U, and $^{232}$Th of materials in the detector and shield systems, radiogenic neutrons in the MeV range are produced in ($\alpha$,n) reactions and spontaneous fission (SF). Additionally, cosmogenic neutrons with energies extending to a few GeV are induced by muons penetrating through the rock into the underground laboratory. This makes the neutron yield dependent on laboratory's depth.

A study of the electronic recoil background in the XENON100 experiment was published in Ref.~\cite{EMBG}. In this paper we summarize results from a comprehensive Monte Carlo study, predicting the neutron induced NR background originating from natural radioactivity and cosmic muons. 
The study of the radiogenic neutron background is based on calculations with the SOURCES-4A code~\cite{SourcesOriginal}. Simulations of the cosmogenic neutron background employ the muon energy spectrum and angular distribution generated with MUSIC and MUSUN~\cite{MUSICandMUSUN}. Neutron production and propagation is performed with the GEANT4 toolkit~\cite{GEANT4}. The detector model developed for the Monte Carlo simulations is described in detail in Refs.~\cite{EMBG} and ~\cite{KishThesis}. 
The results of this work were used to predict the neutron background in the dark matter search data acquired with XENON100 and published in Refs.~\cite{xe100-independent1, xe100-independent2, xe100-run10}. 

\begin{table*}[tbp]
\centering
\caption{Neutron production rates for the materials used in the XENON100 experiment calculated with SOURCES-4A as number of generated neutrons over number of disintegrating U/Th nuclei. The natural abundance of $^{235}$U is taken into account. The systematic uncertainty of the calculation is $\pm$17\%~\cite{SourcesOriginal}. Details on the detector and its components can be found in~Refs.~\cite{xe100-instrument, EMBG}.}
\label{tabDetectorMaterials1}
\vspace{0.2cm}
\begin{tabular}{lclcc}
\hline
\hline
Material 				& Density 			& Chemical composition 										& \multicolumn{2}{c}{Neutron production} \\
		 			& [g/cm$^{3}$] 		&   														&  $^{238}$U (incl. $^{235}$U) 		&  $^{232}$Th	\\
\hline
{\bf Cryostat and TPC} 	& 				& 														&								& \\
316Ti stainless steel		& 8.00 			& C 0.08\%, Si 1\%, Mn 2\%, P 0.045\%, S 0.03\%,					& \multicolumn{2}{c}{see Table~\ref{tabDetectorMaterials2}} \\
					&				& Ni 12\%, Cr 17\%, Mo 2.5\%, Ti 4.0\%, Fe 61.345\%				&								& \\
PTFE				& 2.20 			& CF$_{2}$ 												& (6.3$\pm$1.1)$\times$10$^{-5}$		& (1.0$\pm$0.2)$\times$10$^{-4}$ \\
Copper				& 8.92 			& Cu 100\% 												& (1.1$\pm$0.2)$\times$10$^{-6}$		& (3.6$\pm$0.6)$\times$10$^{-7}$ \\
Ceramics				& 1.00			& NaAlSiO$_{5}$ 											& (1.1$\pm$0.2)$\times$10$^{-5}$		& (2.0$\pm$0.3)$\times$10$^{-5}$ \\
\hline
{\bf PMT parts} 			& 				& 														&								& \\
{\it Kovar} metal		& 8.33 			& Fe 55\%, Ni 29\%, Co 16\%; (13~g/PMT)						& (1.2$\pm$0.2)$\times$10$^{-6}$		& (1.0$\pm$0.2)$\times$10$^{-6}$ \\
Stainless steel 			& 7.64 			& Fe 71.8\%, C 0.1\%, Si 0.5\%, Mn 0.7\%,							& (1.5$\pm$0.3)$\times$10$^{-6}$		& (2.3$\pm$0.4)$\times$10$^{-6}$ \\ 
					&				& Ni 8.6\%, Cr 18.3\%; (7~g/PMT)								&								& \\
Synthetic silica			& 2.20 			& SiO$_{2}$; (2~g/PMT) 										& (2.2$\pm$0.4)$\times$10$^{-6}$		& (2.1$\pm$0.4)$\times$10$^{-6}$ \\
Borosilicate glass		& 2.21 			& SiO$_{2}$ 68.7\%, Al$_{2}$O$_{3}$ 4.3\%, B$_{2}$O$_{3}$ 18.0\%,	& (1.1$\pm$0.2)$\times$10$^{-5}$		& (1.5$\pm$0.3)$\times$10$^{-5}$ \\ 
					& 				&  Li$_{2}$O 1.0\%, Na$_{2}$O 6.0\%, BaO 2.0\%; (1~g/PMT) 			&								& \\
Aluminum				& 2.70 			& Al 100\%; (0.1~g/PMT) 										& (1.4$\pm$0.2)$\times$10$^{-5}$		& (2.8$\pm$0.5)$\times$10$^{-5}$ \\
{\it Cirlex}				& 1.43 	 		& C$_{22}$H$_{10}$N$_{2}$O$_{5}$; (1.4~g/PMT base) 			& (4.8$\pm$0.8)$\times$10$^{-6}$		& (2.4$\pm$0.5)$\times$10$^{-6}$ \\
{\bf Shield components} 	& 				& 														&								& \\
Polyethylene			& 0.92			& CH$_{2}$ 												& (1.9$\pm$0.3)$\times$10$^{-6}$		& (1.4$\pm$0.2)$\times$10$^{-6}$ \\
Lead					& 11.34			& Pb 100\% 												& (1.1$\pm$0.2)$\times$10$^{-6}$		& (3.0$\pm$0.5)$\times$10$^{-11}$ \\
{\bf Environment} 		& 				& 														&								& \\
laboratory concrete		& 2.4				& H 0.89\%, C 7.99\%, O 48.43\%, Na 0.6\%, Mg 0.85\%, 				& (1.9$\pm$0.3)$\times$10$^{-6}$		& (1.5$\pm$0.3)$\times$10$^{-6}$ \\
					& 				& Al 0.9\%, Si 3.86\%, P 0.04\%, S 0.16\%, K 0.54\%, 				&								& \\
					& 				& Ca 34.06\%, Ti 0.04\%, Fe 0.43\%								&								& \\
\hline
\hline
\end{tabular}
\end{table*}

\begin{table*}[tbp]
\centering
\caption{Neutron production due to natural radioactivity in the stainless steel (type 316Ti), mainly used for the XENON100 cryostat and its support bars. The radioactive contamination was measured with gamma- and mass-spectrometry (the most sensitive result of the two methods is given) and shows that secular equilibrium in the chains is broken.}
\label{tabDetectorMaterials2}
\vspace{0.2cm}
\begin{tabular}{lccccc}
\hline
\hline
& \multirow{2}{*}{$^{238}$U$\--$$^{230}$Th}			&  \multirow{2}{*}{$^{226}$Ra$\--$$^{206}$Pb}		&  \multirow{2}{*}{$^{235}$U$\--$$^{207}$Pb}			&  \multirow{2}{*}{$^{232}$Th$\--$$^{228}$Ac}		&  \multirow{2}{*}{$^{228}$Th$\--$$^{208}$Pb}	\\
& & & & & \\
\hline
Contamination [mBq/kg] & 4.9$\pm$1.6  & $<$1.3 & 0.23$\pm$0.06 & $<$0.81 & 2.9$\pm$0.7 \\
\multirow{2}{*}{Neutron production} & \multirow{2}{*}{(1.1$\pm$0.2)$\times$10$^{-6}$}		& \multirow{2}{*}{(3.1$\pm$0.5)$\times$10$^{-7}$}		& \multirow{2}{*}{(4.1$\pm$0.7)$\times$10$^{-7}$}		& \multirow{2}{*}{(1.8$\pm$0.3)$\times$10$^{-9}$}		& \multirow{2}{*}{(2.0$\pm$0.3)$\times$10$^{-6}$} \\
& & & & &  \\
\hline
\hline
\end{tabular}
\end{table*}

\section{Neutron production due to natural radioactivity}
\label{secRadiogenic}

The radiogenic neutron production rates and energy spectra were calculated with the SOURCES-4A software, modified by the group of the University of Sheffield in order to extend the cross sections for ($\alpha$,n) reactions from 6.5~MeV to 10~MeV, based on available experimental data~\cite{SourcesModified}. The calculation was performed with the assumption that the $\alpha$-emitters are uniformly distributed within a homogeneous material.

The systematic uncertainty on the neutron production rate of the SOURCES-4A code is $\pm$17\%~\cite{SourcesOriginal}. A cross-check of the calculations of the neutron production was performed with independent software described in Ref.~\cite{ZhangMeiSoft}, showing agreement in neutron rates to within 20\%.

The program takes into account the energy-dependent ($\alpha$,n) cross sections and {\it Q}-values for all target nuclides, particle stopping cross sections for all elements, the energy of each $\alpha$-particle, and the SF branching ratio for each source nuclide. As an input, the code requires information about source and target nuclides, and the neutron energy range to be considered. The fractions of atoms in the target material were calculated using the chemical composition of the detector and shield components presented in Table~\ref{tabDetectorMaterials1}, and the natural isotopic abundance from Ref.~\cite{NuclearData}. The simulation takes into account all $\alpha$-emitters from each decay chain, and their half-lives, assuming secular equilibrium.

\begin{figure}[!b]
\centering
\includegraphics[width=1.0\linewidth]{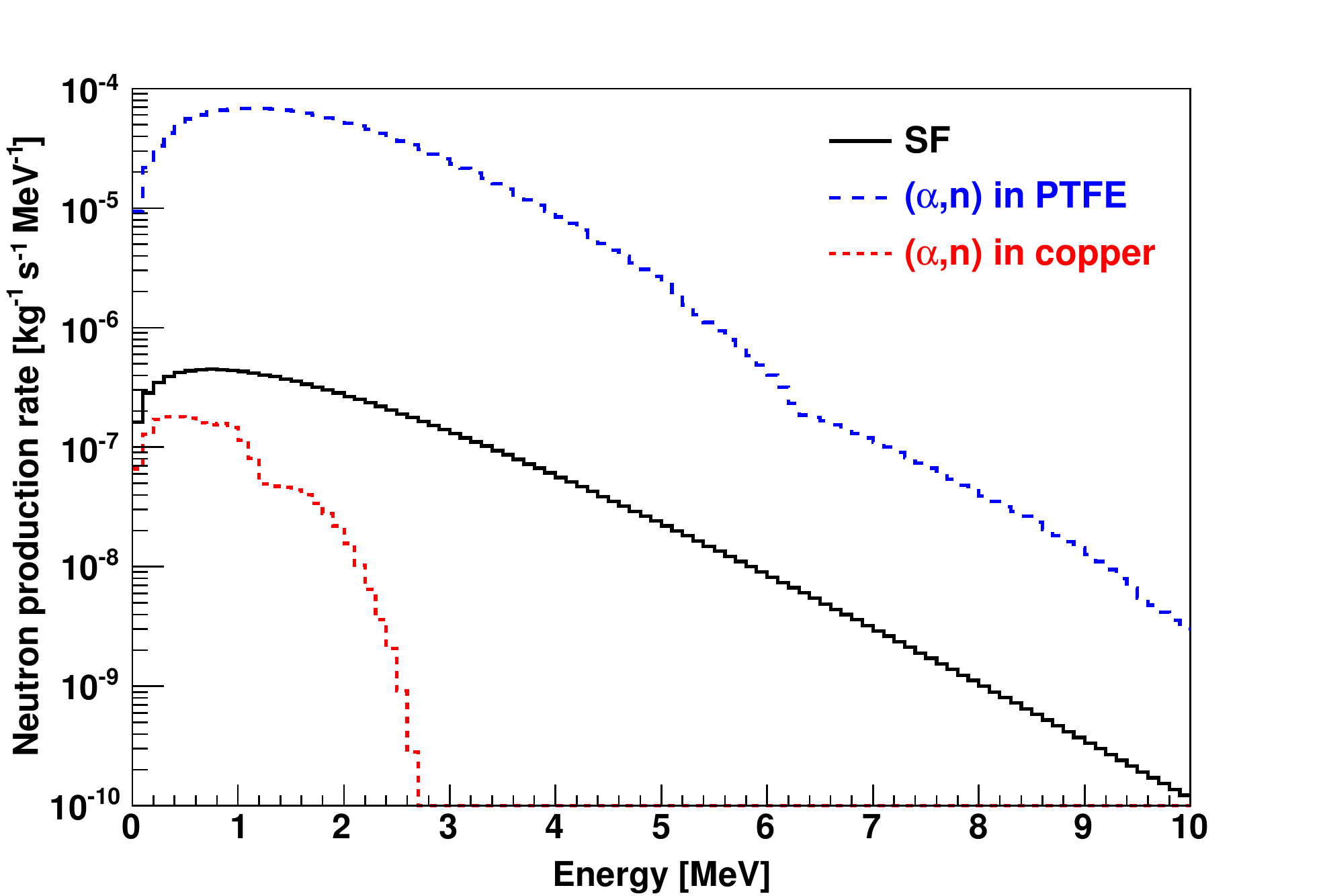}
\caption{Differential neutron production rates in ($\alpha$,n) and spontaneous fission reactions in materials of the XENON100 detector and its shield due to contamination of $^{238}$U, $^{235}$U, and $^{232}$Th (assuming 1~Bq of $^{238}$U and 1~Bq of $^{232}$Th). PTFE is the material with the highest production rate among the XENON100 materials. The neutron production is dominated by ($\alpha$,n) reactions due to the low atomic mass $Z$ of elemental constituents. Copper is an example of a high $Z$ material, where the neutron production rate is almost entirely due to SF reactions. Both, PTFE and copper, have been used to build the TPC field cage in the XENON100 detector.}
\label{figAlphaNproduction1Bq}
\end{figure}

\begin{table*}[tbp]
\centering
\caption{Neutron production rates in the XENON100 components due to ($\alpha$,n) and SF reactions. Some detector components, such as the copper parts inside the cryostat, the TPC resistor chain, the bottom and top electrodes made of 316Ti SS, and PMT cables, are combined into `Additional detector parts' due to their small contribution to the total neutron production. Radioactive contamination of the laboratory concrete is taken from Ref.~\cite{Haffke}, and the neutron production is calculated as a flux.}
\label{tabTotalNeutronProduction}
\vspace{0.2cm}
\begin{tabular}{lcccc}
\hline
\hline
Component  						& Amount  				& \multicolumn{2}{c}{Contamination [mBq/kg]} & Neutron production \\
  		   	 					& 		      				& $^{238}$U & $^{232}$Th &  [neutrons/year] \\
\hline
Cryostat and pumping ports (316Ti SS)	& 73.61~kg 				& \multicolumn{2}{c}{see Table~\ref{tabDetectorMaterials2}} 	 	& 16$\pm$3 \\
Cryostat support bars (316Ti SS)		& 49.68~kg 				& \multicolumn{2}{c}{see Table~\ref{tabDetectorMaterials2}} 		& 13$\pm$2 \\
Detector PTFE 						& 11.86~kg 				& 0.06 					& 0.10 						& 5$\pm$1 \\
PMTs 							& 242~pieces 				& 0.05~/pc				& 0.46~/pc					& 5$\pm$1 \\
PMT bases	 					& 242~pieces 				& 0.16~/pc				& 0.07~/pc					& 12$\pm$2 \\
Additional detector parts				& 						&						&							& 0.20$\pm$0.03 \\
Copper shield						& 2.1$\times$10$^{3}$~kg 	& 0.083 					& 0.012 						& 3.2$\pm$0.5 \\
Polyethylene shield					& 1.6$\times$10$^{3}$~kg 	& 0.23 					& 0.094 						& 37$\pm$6 \\
Lead shield (inner layer)				& 6.6$\times$10$^{3}$~kg 	& 0.66 					& 0.55 						& 162$\pm$28 \\
Lead shield (outer layer)				& 27.2$\times$10$^{3}$~kg 	& 4.20 					& 0.52 						& (4.3$\pm$0.7)$\times$10$^{3}$ \\
LNGS concrete						& 						& 2.6$\times$10$^{4}$		& 8.0$\times$10$^{3}$			& (8.7$\pm$1.5)$\times$10$^{-7}$~cm$^{-2}$s$^{-1}$ \\
\hline
\hline
\end{tabular}
\end{table*}Ä

The cross section for the ($\alpha$,n) reaction is suppressed by the Coulomb barrier for heavy nuclei, and increases with decreasing atomic number $Z$ of the target. 
This explains why the neutron production is dominated by  ($\alpha$,n) reactions for materials consisting of light elements, such as polyethylene (only C and H) or polytetrafluoroethylene (PTFE, C and F).  
For high-$Z$ materials, such as copper and lead, the neutron production is almost entirely due to SF of $^{238}$U (see Fig.~\ref{figAlphaNproduction1Bq}). Thus, the neutron production rate in such materials is dependent only on the contamination level and not on the chemical composition of the material. 

The neutron production rate was calculated as number of generated neutrons over number of disintegrating U/Th nuclei in a given material, with the contamination of $^{235}$U computed from the measured contamination of $^{238}$U, assuming a natural abundance of 0.72\%. The simulated neutron spectra for some of the materials are shown in Fig.~\ref{figAlphaNproduction1Bq}, and the neutron production rates for all materials are presented in Table~\ref{tabDetectorMaterials1}. 

In the 316Ti stainless steel (SS) used in XENON100, the secular equilibrium is broken in the $^{238}$U and $^{232}$Th decay chains. This was established by measuring the intrinsic radioactive contamination by inductively coupled plasma mass spectrometry (ICP-MS), performed in addition to $\gamma$-spectrometry with germanium detectors. Hence, the neutron production in this material was calculated for the different parts of the chains separately: $^{238}$U$\--$$^{230}$Th and $^{226}$Ra$\--$$^{206}$Pb, $^{232}$Th$\--$$^{228}$Ac and $^{228}$Th$\--$$^{208}$Pb, and $^{235}$U$\--$$^{207}$Pb. The results are presented in Table~\ref{tabDetectorMaterials2}. Ignoring the disequilibrium, the neutron background would be underestimated.

The total neutron production rates were calculated by scaling the results of SOURCES-4A to the mass of the components in the detector and shield and to their measured radioactive contamination, using the mass model and the radioactive screening results introduced in Refs.~\cite{EMBG, ScreeningPaper}. The results are presented in Table~\ref{tabTotalNeutronProduction}.

The detector components with the highest total neutron production rates are the lead and polyethylene in the shield, and the detector cryostat and support bars made from 316Ti SS. Neutron production in the TPC resistor chain is negligible, due to the small mass involved. 
Even though the neutron production rate in the aluminum of the PMTs is relatively high, its contribution to the total neutron production in the PMTs is negligible due to the very low amount of material (0.1~g per PMT), since it is used only as strips deposited on the window in order to improve the resistivity of the photocathode at cryogenic temperature.

Neutron production due to natural radioactivity in the concrete walls of LNGS was calculated using the measured chemical composition of Ref.~\cite{Wulandari}. Radiogenic production in the rock has been ignored in the present study, since our simulation showed that almost all neutrons which originate in the rock are absorbed by the concrete shell. The results of our simulations agree well with the measurements of neutron flux  summarized in the same paper. In particular, the discrepancy with the values measured by~\cite{NeutronFlux1, NeutronFlux2} is less than 15\%.

The neutron energy spectra calculated from SOURCES-4A and the total production rates were used as an input for Monte Carlo simulations to predict the neutron-induced NR background in the XENON100 experiment. 
The neutron propagation was performed with GEANT4.9.3.p02, using the neutron data files G4NDL~3.13 with thermal cross sections, which are based on the ENDF/B-VI/B-VII databases~\cite{G4NDL}. For each material and neutron source, 1 million events were simulated, resulting in a negligible statistical uncertainty of $\sim$1\%.

In the analysis of the simulated data, only `true' NRs in the sensitive volume were selected, meaning that all events containing an ER component have been discarded. Such a cut has a residual error, as it erroneously removes events where a Xe nucleus is excited by an inelastic neutron scatter to a relatively long-lived state, e.g. $^{\textrm{129m}}$Xe with T$_{1/2}$~=~8.9~days, and $^{\textrm{131m}}$Xe with T$_{1/2}$~=~11.8~days. These events have a signature of a prompt NR, followed by an ER produced by a $\gamma$ from de-excitation of the metastable state.  However, by computing the ratio of the cross sections for elastic and inelastic neutron interactions with xenon nuclei, restricted to these particular nuclear levels, we estimated that the contribution of these events to the total NR rate is $<$2\%, therefore irrelevant for the XENON100 background prediction.

\begin{figure}[!tbp]
\includegraphics[width=1.0\linewidth]{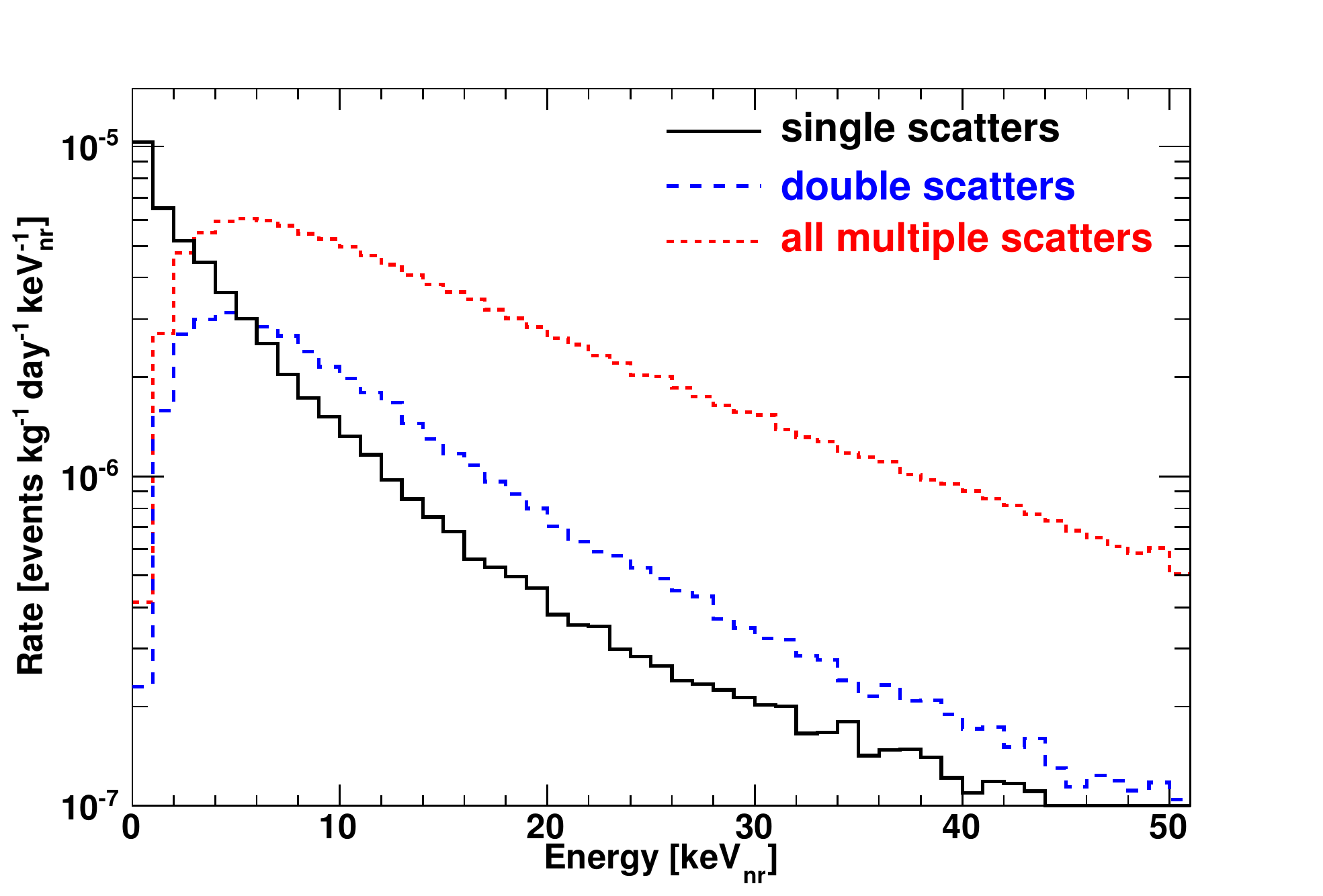}
\caption{Energy spectra of NRs in 34~kg fiducial volume from neutrons produced in ($\alpha$,n) and spontaneous fission reactions. The total energy deposited in multiple scatter interactions is on average higher than that of single scatter events.}
\label{figSpectraSinglesMultiples}
\end{figure}

Single and multiple scatters are distinguished by taking into account the ability of the XENON100 detector to separate interactions close in Z direction. It is limited to two scatters separated by more than $\sim$3~mm~\cite{xe100-instrument, EMBG} in Z, due to the diffusion of the electron cloud and the gas gap, which define the typical width of the proportional scintillation (S2) signal. The energy spectra of NRs produced in single and multiple scatter neutron interactions are shown in Fig.~\ref{figSpectraSinglesMultiples} for the 34~kg fiducial mass used in Ref.~\cite{xe100-run10}. 

The energy range for the WIMP search performed with 100.9~days exposure in 2011~\cite{xe100-independent2} was 4 to 30 photoelectrons (PE). The lower threshold has been set to 3~PE in the 2012 analysis~\cite{xe100-run10} of the 224.6~days exposure, as the acceptance was still high at this value, and due to the improved electronic noise conditions. The upper threshold was kept at 30~PE as the sensitivity does not increase significantly at higher energies. The NR energy scale is derived from the S1 signal using the relative scintillation efficiency ${\cal L}_\mathrm{eff}$ described in Ref.~\cite{xe100-independent2}. Due to a change in the measured value of the LXe response to 122~keV$_\mathrm{ee}$ gamma rays, used to normalize the scale, the energy ranges used for the 2011 and 2012 results are slightly different, and correspond to (8.4$\--$44.6)~keV$_{\mathrm{nr}}$ and (6.6$\--$43.3)~keV$_{\mathrm{nr}}$, respectively.

\begin{figure}[!t]
\includegraphics[width=1.0\linewidth]{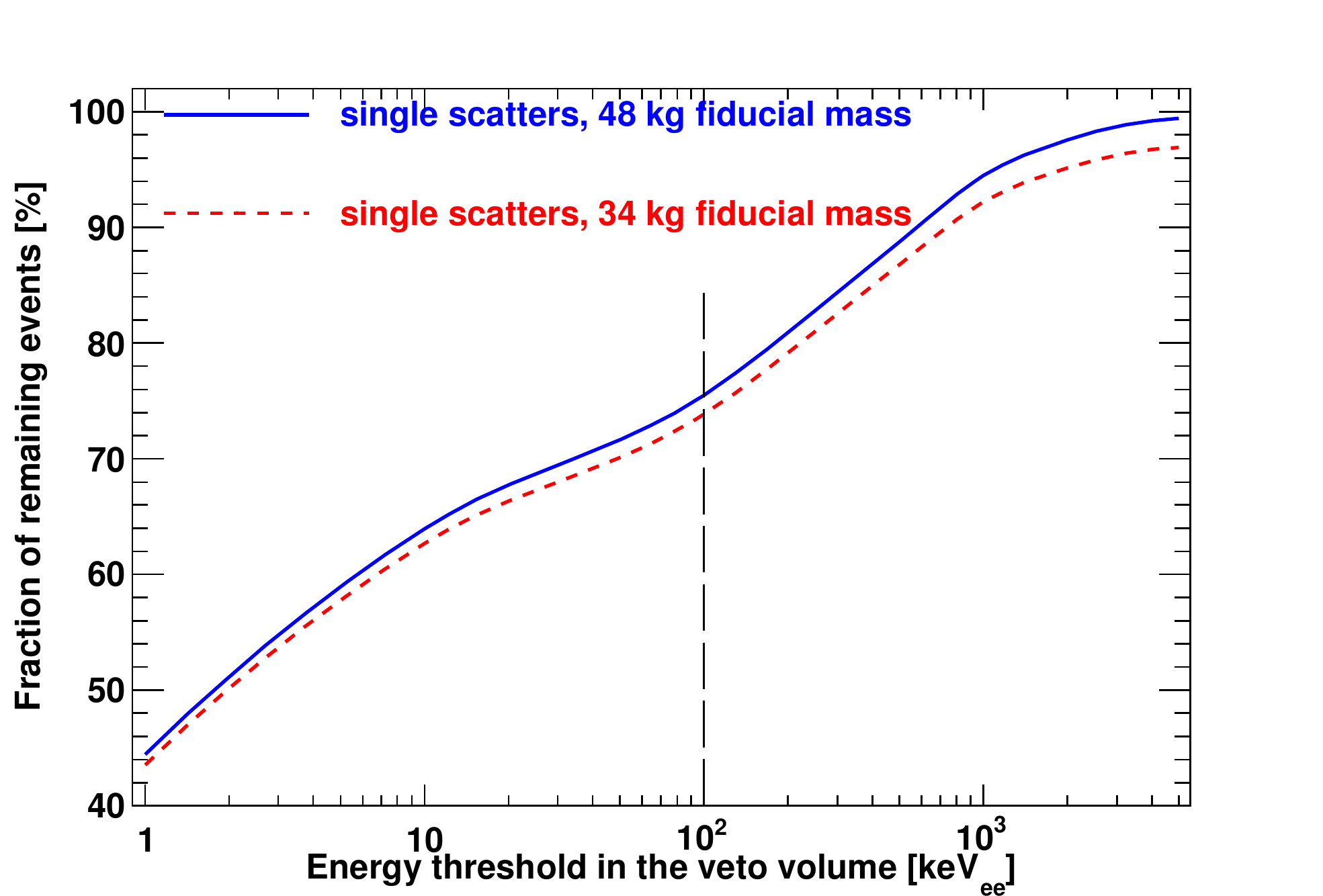}
\caption{Efficiency of the veto coincidence cut as a function of the energy threshold in the veto volume. A veto cut with the measured volume-averaged energy threshold of 100~keV$_\mathrm{ee}$ (shown as a vertical dashed line) provides a $\sim$25\% reduction of the single scatter NR rate.}
\label{figRateVetoThreshold}
\end{figure}

\begin{figure}[!tbp]
\centering
\includegraphics[width=1.0\linewidth]{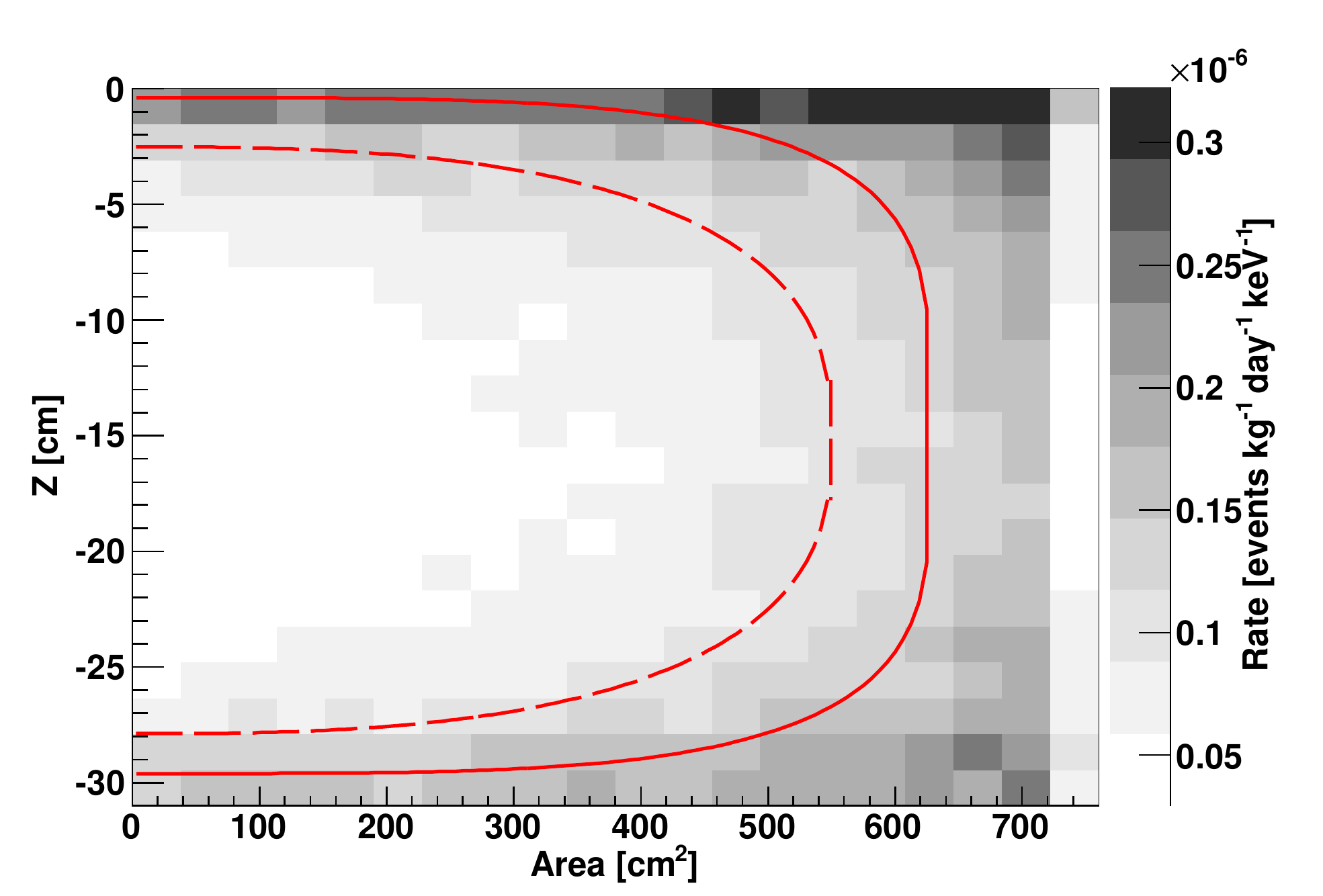}
\caption{Spatial distribution of single scatter NRs produced by radiogenic neutrons in the energy range of interest for the WIMP search. 
The solid (dashed) line shows the 48~kg (34~kg) fiducial volume boundary. The lower density at the edge is due to the specific shape of the TPC defined by interlocking PTFE panels. This leads to a smaller active LXe volume represented by the last radial bins.}
\label{figAZalphan}
\end{figure}

The spatial distribution of single scatter NRs in the energy region of interest for a WIMP search is shown in Fig.~\ref{figAZalphan}. The radiogenic NR background was predicted for the entire 62~kg target volume and for two fiducial volumes, 48~kg and 34~kg, which were used in the analyses published in 2011~\cite{xe100-independent2} and 2012~\cite{xe100-run10}, respectively. Fiducialization is less efficient for reducing the NR background because of the longer mean free path of neutrons ($\sim$14~cm for 1~MeV), compared to $\gamma$-rays of similar energy ($\sim$6~cm).

The background rate of single scatter NRs in the target volume can be reduced by applying a veto coincidence cut~\cite{xe100-analysis}, which rejects events where a particle deposits energy in the active LXe veto which surrounds the TPC. This can be either due to the same neutron which produces a NR in the target or due to an associated prompt $\gamma$-ray, for example produced by an inelastic neutron interaction. ERs and NRs in the veto volume cannot be distinguished through the ratio of scintillation and ionization signals, as it is done within the TPC. Hence, energy depositions from all interactions in the veto volume are summed up, taking into account the light quenching for NRs~\cite{xe100-independent2}. As shown in Fig.~\ref{figRateVetoThreshold}, by applying a cut with the measured volume-averaged energy threshold of 100~keV$_{\mathrm{ee}}$~\cite{xe100-instrument, EMBG}, the NR background can be reduced by $\sim$25\% with respect to the `passive veto' mode, when only self-shielding of the external LXe layer is taken into account.

\begin{table*}[!t]
\centering
\caption{Predicted NR background rate in the given energy range from neutrons produced in ($\alpha$,n) and SF reactions due to natural radioactivity in the detector and shield components. The statistical error of the GEANT4 simulation is $\sim$1\%, hence the total error is dominated by the systematic uncertainty of the calculation with SOURCES-4A. The active veto coincidence cut assumes an average detection threshold of 100~keV$_{\mathrm{ee}}$ in the LXe active veto. Only single scatter NRs are relevant as background for WIMP searches. The acceptance to NRs, while relevant for the WIMP search, is not taken into account here, but in Section~\ref{secBackgroundPredictions}. The multiplicity of neutron interactions is indicated with the parameter {\it n}.}
\label{tabAlphaNrates}
\begin{tabular}{lccccccc}
\hline
\hline
				      					& & \multicolumn{6}{c}{Predicted background rate [year$^{-1}$]} \\
\multicolumn{2}{l}{Target volume} 			& \multicolumn{2}{c}{62~kg} 				& \multicolumn{2}{c}{48~kg}  				& \multicolumn{2}{c}{34~kg} \\
\multicolumn{2}{l}{Energy range}			& \multicolumn{2}{c}{8.4$\--$44.6~keV$_{\mathrm{nr}}$}		& \multicolumn{2}{c}{8.4$\--$44.6~keV$_{\mathrm{nr}}$}			& \multicolumn{2}{c}{6.6$\--$43.3~keV$_{\mathrm{nr}}$} \\
\multicolumn{2}{l}{Veto}					& passive			& active				& passive			& active				& passive		& active \\
\hline
single scatter events 	& ({\it n}=1)		& 0.49$\pm$0.08	& 0.38$\pm$0.07		& 0.28$\pm$0.05 	& 0.22$\pm$0.04		& 0.18$\pm$0.03 	& 0.14$\pm$0.02	\\
double scatter events 	& ({\it n}=2)		& 0.46$\pm$0.07 	& 0.34$\pm$0.06		& 0.34$\pm$0.06 	& 0.25$\pm$0.04		& 0.25$\pm$0.04 	& 0.19$\pm$0.03	\\
multiple scatter events 	& ({\it n}$>$1)		& 1.19$\pm$0.20 	& 0.85$\pm$0.15		& 0.93$\pm$0.16 	& 0.66$\pm$0.11		& 0.74$\pm$0.13 	& 0.54$\pm$0.09	\\
all events 				& ({\it n}$>$0)		& 1.69$\pm$0.29 	& 1.23$\pm$0.21		& 1.21$\pm$0.21 	& 0.88$\pm$0.15		& 0.92$\pm$0.16 	& 0.68$\pm$0.12	\\
\hline
\hline
\end{tabular}
\end{table*}

\begin{table}[!h]
\centering
\caption{Relative contribution of different components to the total single scatter NR background due to radiogenic neutrons in XENON100. The fractions do not significantly change with the fiducialization of the target volume. `Other components' include: polyethylene and lead shield, copper parts of the TPC, 316Ti SS support rings for the mesh electrodes, TPC resistor chain, PMT cables, and the concrete of the laboratory.}
\label{tabPieAlphaN}
\vspace{0.2cm}
\begin{tabular}{lcc}
\hline
Component      						& Contribution [\%] \\
\hline
Cryostat and pumping ports (316Ti SS) 	& 26		\\
Detector PTFE 						& 22		\\
PMTs 							& 21		\\
Cryostat support bars (316Ti SS) 		& 14		\\
PMT bases 						& 8		\\
Copper shield 						& 5		\\
Other components 					& 4		\\
\hline
\hline
\end{tabular}
\end{table}

The predicted NR background rates in the WIMP search region from radiogenic neutrons are presented in Table~\ref{tabAlphaNrates}. The contribution of the different components to the total background is shown in Table~\ref{tabPieAlphaN}. The relative contributions do not significantly change by applying fiducial volume and veto coincidence cuts. The dominant part of the background comes from the 316Ti SS components, PMTs, and the PTFE parts of the TPC. This was expected due to the rather high neutron production rates in these components and their location close to the liquid xenon target. Despite the high neutron production rates in the lead shield, the contribution of this source to the total NR background rate is negligible, since it is located outside of the polyethylene neutron shield. 
The contribution from radiogenic neutrons originating from the concrete lining of the laboratory's cavern results in (6.6$\pm$0.4)$\times$10$^{-3}$~events/year in the entire target volume (62~kg of LXe) in the energy range of interest for the WIMP search, even without using a veto coincidence cut, and hence can be considered negligible.

\section{muon-induced neutron production}
\label{secCosmogenic}

\begin{figure}[!b]
\includegraphics[width=1.0\linewidth]{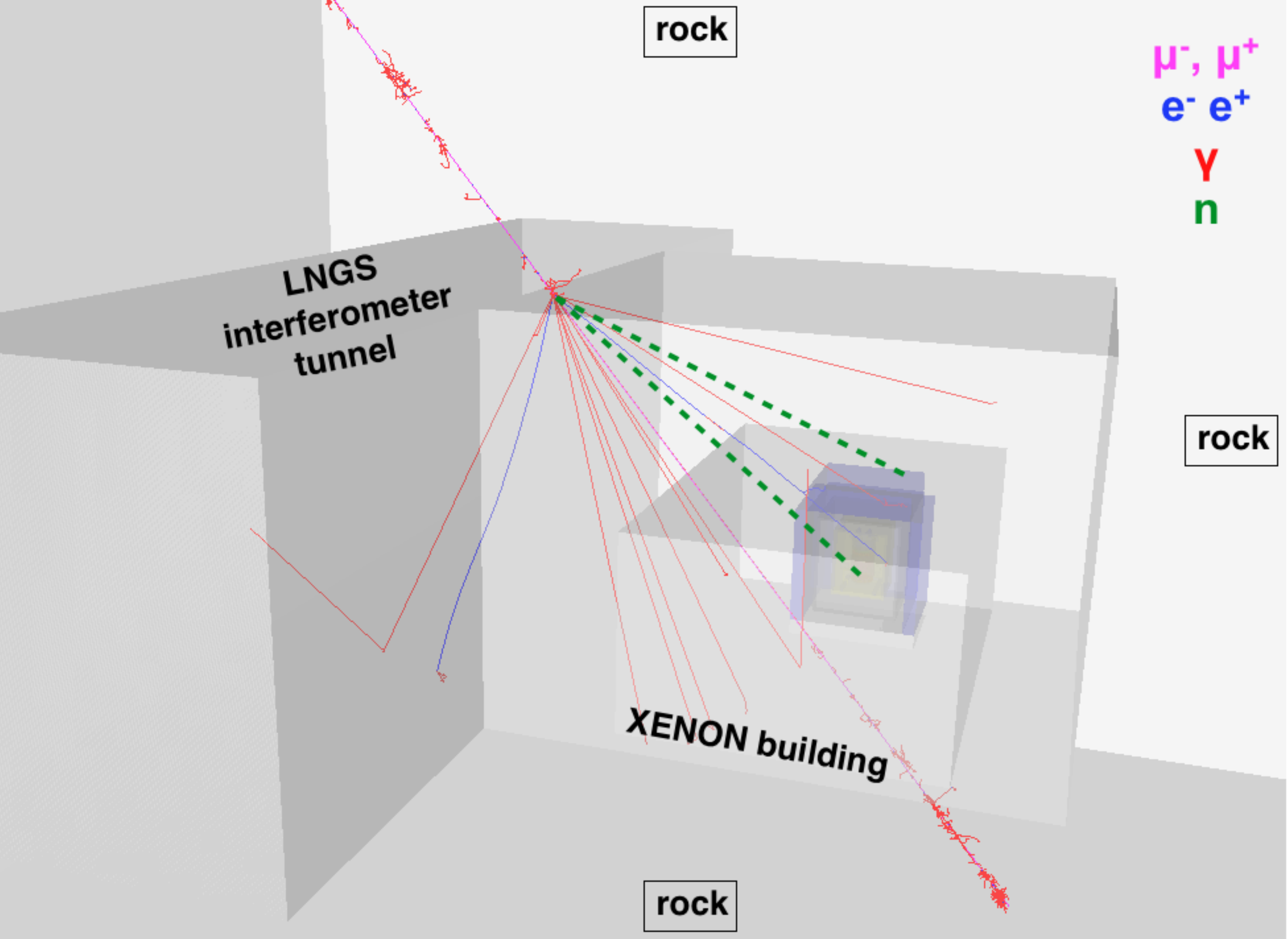}
\caption{The GEANT4 model of the XENON100 experimental site for simulations of the cosmogenic neutron background, showing an example of a muon interaction. Two neutrons (dashed lines) and an electromagnetic shower are produced in the rock: one neutron is stopped by the shield, and another one penetrates into the detector volume.}
\label{figModelScreenshot}
\end{figure}

The cosmic muon flux underground at LNGS is reduced by six orders of magnitude with respect to the value measured at the surface, due to the 3.6~km-water-equivalent, obtained averaging over the muon arrival direction, of overburden rock~\cite{LVD_DepthIntensity}. 
High-energy muons penetrating into the underground laboratory produce neutrons through photo-nuclear reactions in electromagnetic showers, in deep inelastic muon-nucleus interactions, and in several secondary processes ($\pi$-n, $\pi$-absorption, p-n, etc.)~\cite{MuonInducedNeutronProduction_1, MuonInducedNeutronProduction_2}. The deeper the experimental site, the higher the mean muon energy and hence the average energy of muon-induced neutrons. Moreover, neutron production due to negative muon capture, which is relevant for low energy stopping muons, becomes negligible. The energy of muon-induced neutrons extends up to a few GeV, hence the hydrocarbon and water neutron shield, as employed in XENON100, cannot fully moderate and capture them.

In order to simulate the muon-induced neutron background, the GEANT4 model introduced in Ref.~\cite{EMBG} has been extended by including the rock and concrete lining of the underground site at LNGS, taking into account a rock thickness of 5~m. 
The Gran Sasso rock is composed mainly of CaCO$_{3}$ and MgCO$_{3}$, and has an average density of (2.71$\pm$0.05)~g/cm$^{3}$~\cite{LNGSrock}. The location of the experiment, situated in a cavity at the corner of the interferometer tunnel at LNGS, has been described with a simplified geometry. The model is shown in Fig.~\ref{figModelScreenshot}, together with an example of a muon interaction in the rock of the laboratory, which generates two neutrons and an electromagnetic shower: one of the neutrons is stopped by  the passive shield, while the other one penetrates into the target volume.

\begin{figure}[!ht]
\centering
\subfigure[~~energy spectrum]{
\includegraphics[width=1.0\linewidth]{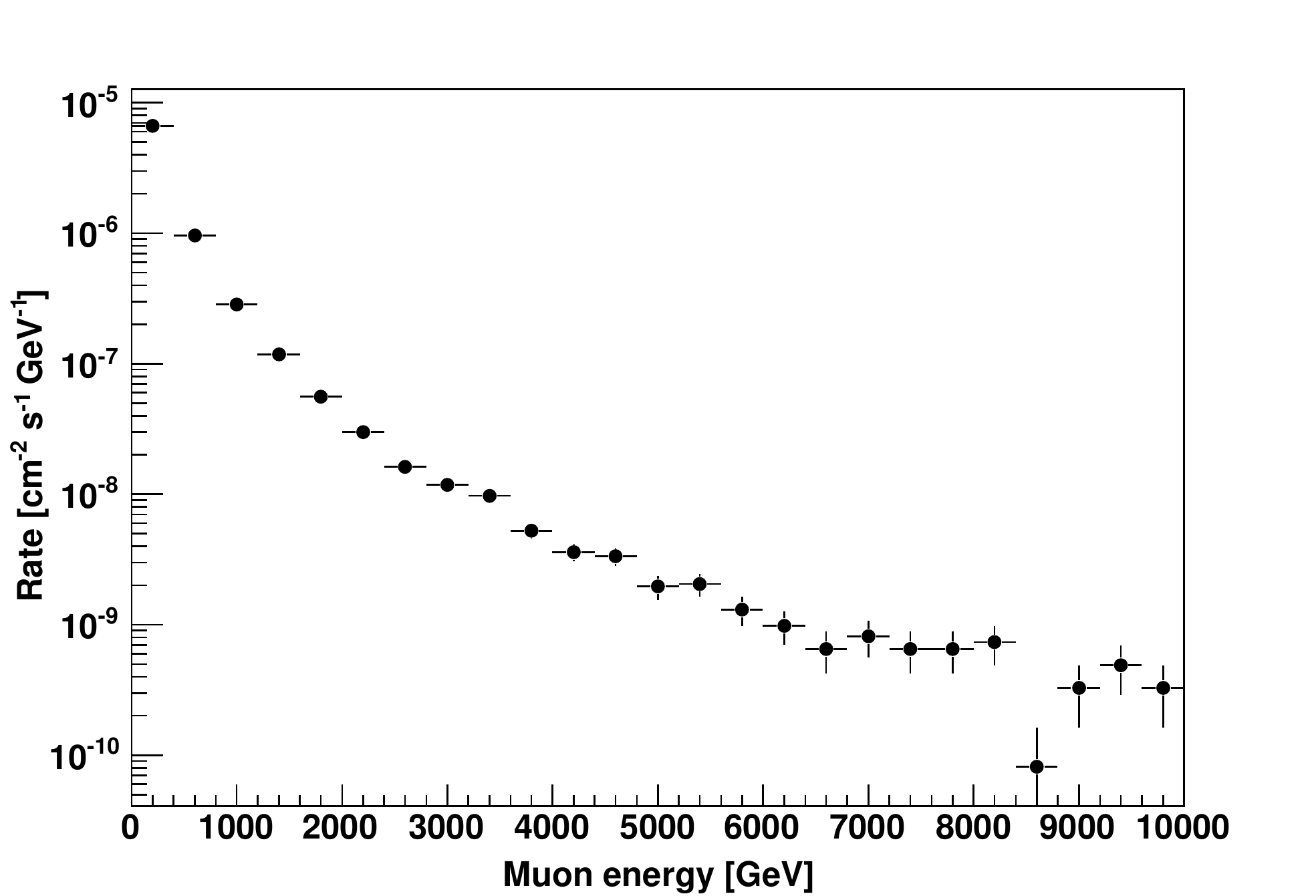}
\label{figMuonAngularSpectrum_1}}
\subfigure[~~angular spectrum]{
\includegraphics[width=1.0\linewidth]{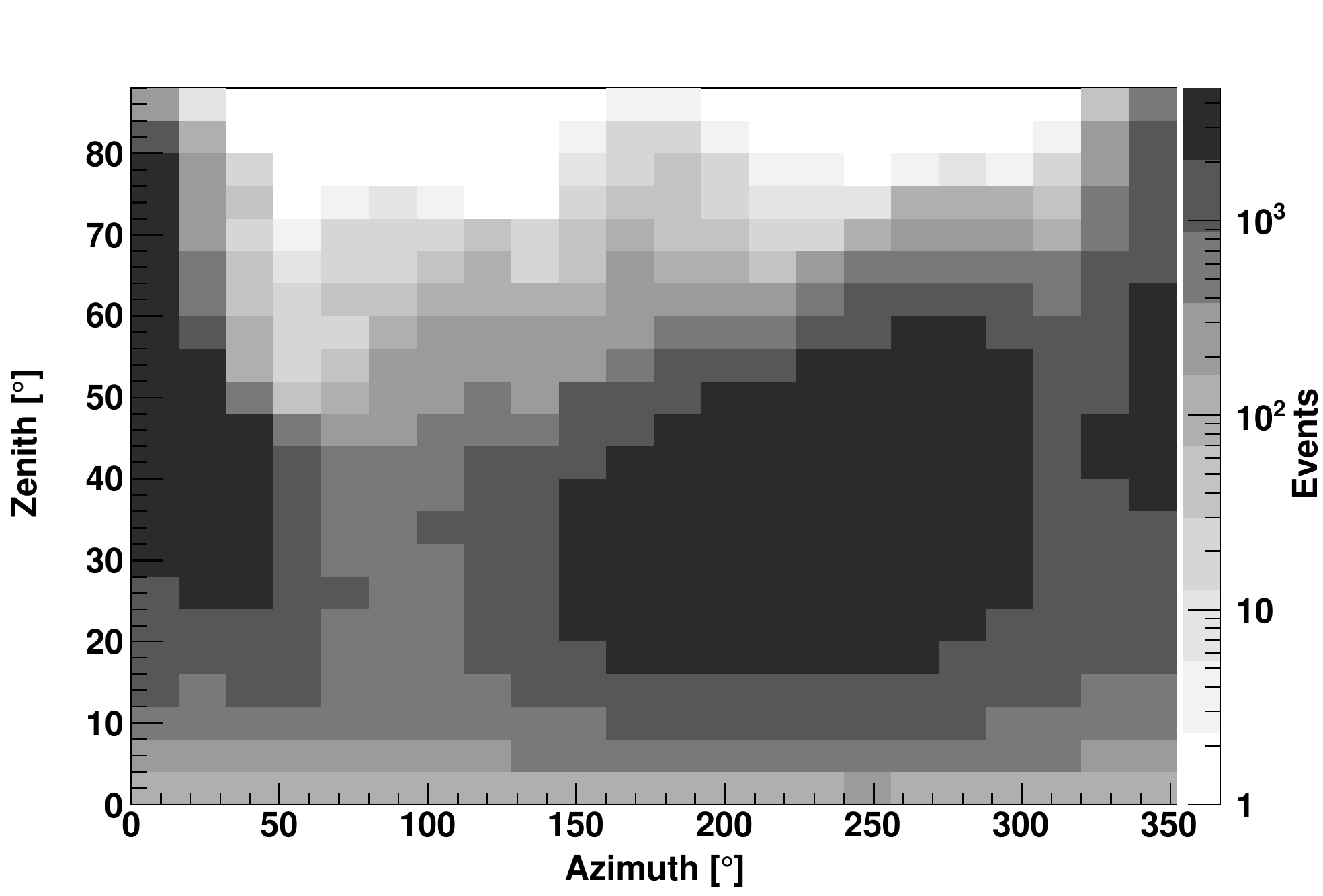}
\label{figMuonAngularSpectrum_2}}
\caption{Energy and angular spectra of the muons underground at LNGS from simulations with MUSUN. The average muon energy is 273~GeV, and most of the muons have a zenith angle $<$60$^{\circ}$. }
\label{figMuonAngularSpectrum}
\end{figure}

An average muon flux of 1.2 h$^{-1}$m$^{-2}$ was assumed for the simulations, as measured by MACRO~\cite{MuonIntensity_MACRO} and LVD~\cite{MuonIntensity_LVD}. The seasonal variation of the muon intensity~\cite{MuonSeasonalModulation}, being smaller than 2\%, is neglected. 
The muon energy spectrum and angular distribution were simulated with the MUSIC-MUSUN package~\cite{MUSICandMUSUN}, taking into account the depth of the experimental hall and the details of the mountain profile; the results are shown in Fig.~\ref{figMuonAngularSpectrum}. The average muon energy is 273~GeV, which is in good agreement with the results in Ref.~\cite{MeiHime} and the measurements reported in Ref.~\cite{MuonIntensity_MACRO}. The azimuthal and zenith distributions agree well with the measurement of the LVD experiment~\cite{LVD_DepthIntensity, MuonIntensity_LVD}; there is agreement also with the predictions done with FLUKA and the measurements by MACRO and Borexino~\cite{AngleFLUKA}. Most muon trajectories have zenith angles $<$60$^{\circ}$. The $\mu^{+}$/$\mu^{-}$ ratio is assumed to be 1.4, as shown by recent observations for high energy muons~\cite{MuonRatio1, MuonRatio2}. 

The propagation of the high energy muons was performed  with GEANT4.9.3.p02 using the QGSP~BIC~HP physics list \cite{G4_PhysicsList}, which is based on a quark gluon string model for high energy hadronic interactions~\cite{QGSP}, and with a data-driven high-precision neutron package to transport neutrons below 20~MeV down to thermal energies. For primary protons and neutrons with energies below 10~GeV, the GEANT4 binary cascade is used, which describes production of secondary particles in interactions with nuclei~\cite{G4_PhysicsList} more accurately than other GEANT4 models. 
The direct interaction between muons and nuclei is modeled with the G4MuNuclearInteraction process~\cite{G4MuNuclInteraction}, which describes it by producing virtual photons and treating them as a combination of $\pi^{+}$ and $\pi^{-}$ interactions.

About 300 million muons were simulated, corresponding to about 185~years of livetime. This results in a statistical uncertainty of $\sim$10\% on the background prediction.
Information from the literature was used in order to evaluate the systematic uncertainty of the simulations. 
The muon-induced neutron production with GEANT4 was validated in Ref.~\cite{MuonNeutronProd_NA55} via comparison with measured data from the NA55 experiment, resulting in a factor of $\sim$2 underproduction by the Monte Carlo simulation. 
Such comparison has been also performed in Ref.~\cite{MuonNeutronProd_ZeplinAraujo, MuonNeutronProd_Zeplin}, using the experimental data of ZEPLIN-II, and in Ref.~\cite{KamLAND} for the KamLAND experiment. Both studies indicate a factor of $\sim$2 overproduction for heavy targets, such as lead, by GEANT4. These results have been used to set the systematic uncertainty of the simulations for GEANT4 by assigning asymmetric error bars, which are the size of a factor $\pm$2 relative to the mean value. At the same time, a new study performed by ZEPLIN-III~\cite{zeplin3muons} shows that more accurate results can be achieved using updated versions of GEANT4 and physics models; in this case Monte Carlo simulations result in a higher neutron production yield, which in a better agreement with the measured data. Based on these results, and taking into account the used GEANT4 versions, we expect the difference within the systematic uncertainty assumed in our study.

\begin{table*}[!t]
\centering
\caption{Predicted NR background rate due to muon-induced neutrons. The active veto coincidence cut assumes a volume-averaged energy threshold of 100~keV$_{\mathrm{ee}}$. The statistical error of the GEANT4 simulation is $\sim$10\%. A factor 2 systematic uncertainty in neutron production rates in GEANT4 is assumed from the comparison of simulations and measured data from the NA55~\cite{MuonNeutronProd_NA55} and ZEPLIN-II~\cite{MuonNeutronProd_ZeplinAraujo, MuonNeutronProd_Zeplin} experiments. The multiplicity of neutron interactions is indicated with the parameter {\it n}. No deficit of acceptance for NRs is considered yet (see Section~\ref{secBackgroundPredictions}).}
\label{tabMuonNrates}
\begin{tabular}{lcccccccc}
\hline
\hline
				      				& & \multicolumn{6}{c}{Predicted background rate [year$^{-1}$]} \\
\multicolumn{2}{l}{Target volume} 		& \multicolumn{2}{c}{62~kg} 							& \multicolumn{2}{c}{48~kg}  							& \multicolumn{2}{c}{34~kg} \\
\multicolumn{2}{l}{Energy range}		& \multicolumn{2}{c}{8.4$\--$44.6~keV$_{\mathrm{nr}}$}		& \multicolumn{2}{c}{8.4$\--$44.6~keV$_{\mathrm{nr}}$}		& \multicolumn{2}{c}{6.6$\--$43.3~keV$_{\mathrm{nr}}$} \\
\multicolumn{2}{l}{Veto}				& passive					& active					& passive					& active					& passive					& active \\
\hline
\multirow{2}{*}{single scatter events} 	& \multirow{2}{*}{({\it n}=1)}		& \multirow{2}{*}{2.2$^{+2.2}_{-1.1}$}		& \multirow{2}{*}{0.9$^{+0.9}_{-0.5}$}		& \multirow{2}{*}{1.3$^{+1.3}_{-0.6}$} 		& \multirow{2}{*}{0.5$^{+0.5}_{-0.3}$}		& \multirow{2}{*}{0.9$^{+0.9}_{-0.5}$} 	& \multirow{2}{*}{0.3$^{+0.3}_{-0.2}$}	\\
& & \\
double scatter events 				& ({\it n}=2)					& 1.8$^{+1.8}_{-0.9}$ 					& 0.7$^{+0.7}_{-0.4}$					& 1.2$^{+1.2}_{-0.6}$ 						& 0.5$^{+0.5}_{-0.3}$					& 1.0$^{+1.0}_{-0.5}$ 					& 0.3$^{+0.3}_{-0.2}$				\\
\multirow{2}{*}{multiple scatter events} 	& \multirow{2}{*}{({\it n}$>$1)}		& \multirow{2}{*}{5.6$^{+5.6}_{-2.8}$} 	& \multirow{2}{*}{2.0$^{+2.0}_{-1.0}$}		& \multirow{2}{*}{4.3$^{+4.3}_{-2.2}$} 		& \multirow{2}{*}{1.5$^{+1.5}_{-0.8}$}		& \multirow{2}{*}{3.4$^{+3.4}_{-1.7}$} 	& \multirow{2}{*}{1.0$^{+1.0}_{-0.5}$}	\\
& & \\
all events 							& ({\it n}$>$0)					& 7.8$^{+7.8}_{-3.9}$					& 2.9$^{+2.9}_{-1.4}$					& 5.6$^{+5.6}_{-2.8}$ 						& 2.1$^{+2.1}_{-1.0}$					& 4.3$^{+4.3}_{-2.2}$ 					& 1.4$^{+1.4}_{-0.7}$				\\
\hline
\hline
\end{tabular}
\end{table*}

The background rates from muon-induced neutrons are presented in Table~\ref{tabMuonNrates}, and the contribution of different detector components to the total muon-induced neutron background is given in Table~\ref{tabNeutronProductionMuons}. It was calculated for all neutrons that produce NRs in the target volume, and also for those neutrons that have a true single scatter NR signature and contribute to the background in the WIMP search energy region of interest (ROI).  
The production rate of neutrons in liquid xenon is relatively high. However, they do not contribute significantly to the NR background, as they are coincident with high energy depositions from the primary muon or an associated electromagnetic cascade. The largest contribution (55\%) to the background is from neutrons generated in the innermost (copper) shield.

\begin{table}[!h]
\centering
\caption{Relative contribution from the detector and shield components to the muon-induced neutron background. It is given for all neutrons that produce NRs  in the target volume (left column), and also for those neutrons that have only a true single scatter NR signature in the energy region of interest (ROI). }
\label{tabNeutronProductionMuons}
\vspace{0.2cm}
\begin{tabular}{lcc}
\hline
\hline
Component/Material      		& \multicolumn{2}{c}{Contribution [\%]} \\
  		     		      		& all NRs~~~~~~~~ 	& single scatter \\
  		     		      		&  				& NRs in ROI  \\
\hline
Rock and concrete 			&  $<$1 			&  5		\\
Water shield 				&  $<$1			&  5		\\
Lead shield 				&  6 				&  15		\\
Polyethylene shield 			&  2				&  5		\\
Copper shield 				&  33				&  55		\\
Cryostat and detector ports	&  3				& $<$1		\\
Detector PTFE 				&  5				&  10 	\\
LXe 						&  46				&  5		\\
Other components			&  5				& $<$1		\\
\hline
\hline
\end{tabular}
\end{table}

The spatial distribution of NRs due to cosmogenic neutrons is more uniform than those of radiogenic origin, due to their higher energies and therefore longer mean free path in the LXe. Therefore, fiducialization of the target volume is not efficient in reducing this background source. As detectors get larger, identification of these neutrons becomes easier through their detected multiplicity.

\begin{figure}[!tbp]
\includegraphics[width=1.0\linewidth]{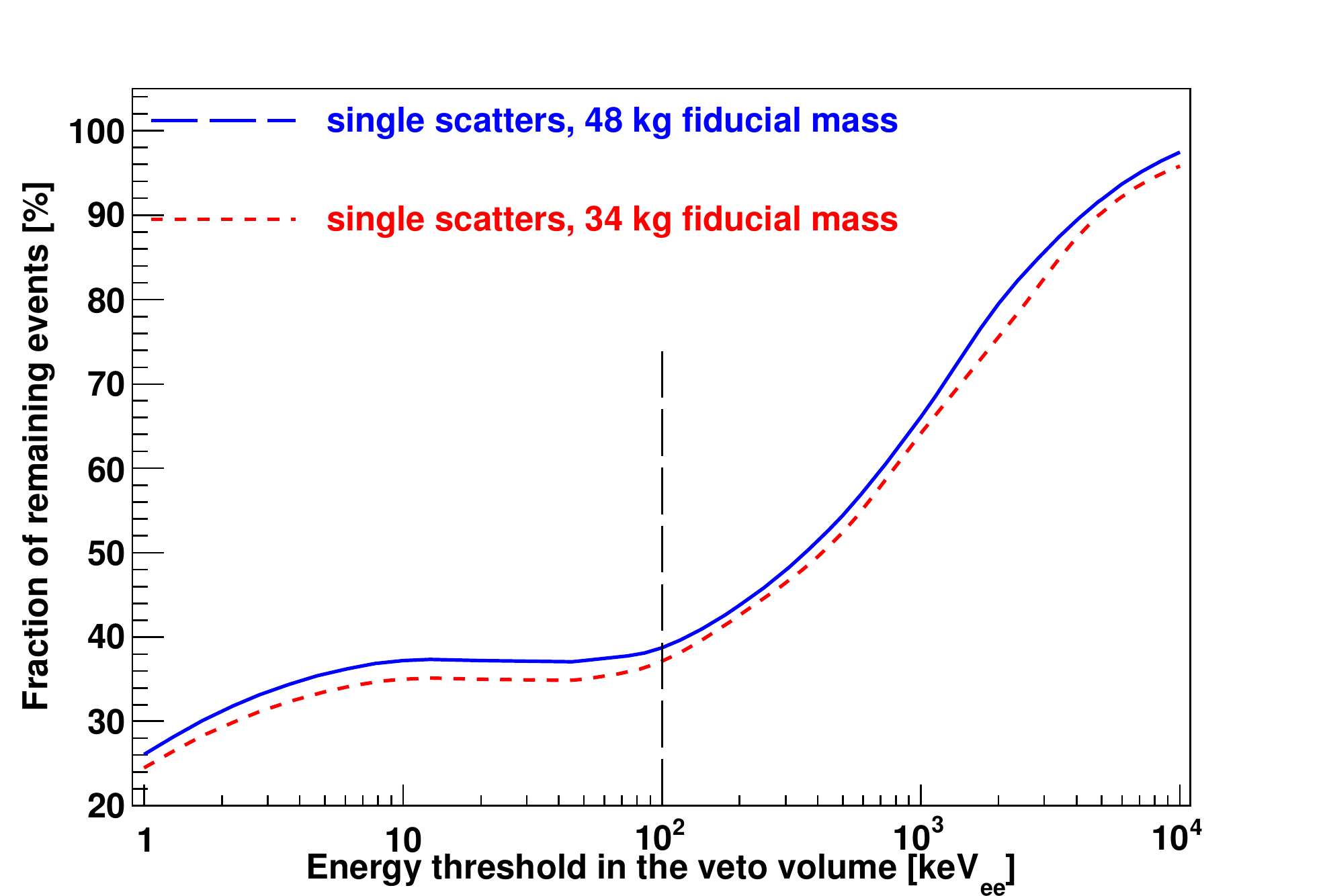}

\caption{Efficiency of the veto coincidence cut as a function of the energy deposited in the veto volume for reduction of the cosmogenic neutron background. A veto cut with the measured volume-averaged energy threshold of 100~keV$_{\mathrm{ee}}$ provides a $\sim$60\% reduction of the single scatter NR rate.}
\label{figRateVetoThresholdMuons}
\end{figure}

Since muon-induced neutron production is often accompanied by an electromagnetic cascade and by a high energy deposition from the incident muon, this background can be reduced by applying a veto coincidence cut. As shown in Fig.~\ref{figRateVetoThresholdMuons}, a veto coincidence cut with the measured volume-averaged threshold of 100~keV$_{\mathrm{ee}}$ rejects more than half of the single scatter NRs in the target volume.

\section{Total predicted nuclear recoil background}
\label{secBackgroundPredictions}

The Monte Carlo simulations of the NR background presented above assume 100\% detection efficiency. The measured trigger efficiency is $>$99\% at 300 (150)~PE in the 2011 (2012) WIMP search results and rolls off at lower values~\cite{xe100-instrument}, which reduces the detection efficiency. In addition, the acceptance for single scatter events is finite, and is determined by the size of the proportional scintillation (S2) signal. If one interaction of a double scatter event generates an S2 which is below the threshold, such an event is mis-identified as a single scatter interaction. Since in the measured data the energy of an event is computed from the scintillation signal, the average energy of these events is higher than that of true single scatter interactions. 
Electronic recoils are less affected by this detection efficiency, as the typical S2 signals are much larger than the threshold value. In order to select single scatter NR interactions in the Monte Carlo data, we apply cuts which are similar to the ones used in the analysis of the measured data.

\begin{figure}[!hb]
\centering
\subfigure[~~raw Monte Carlo data]{
\includegraphics[width=1.0\linewidth]{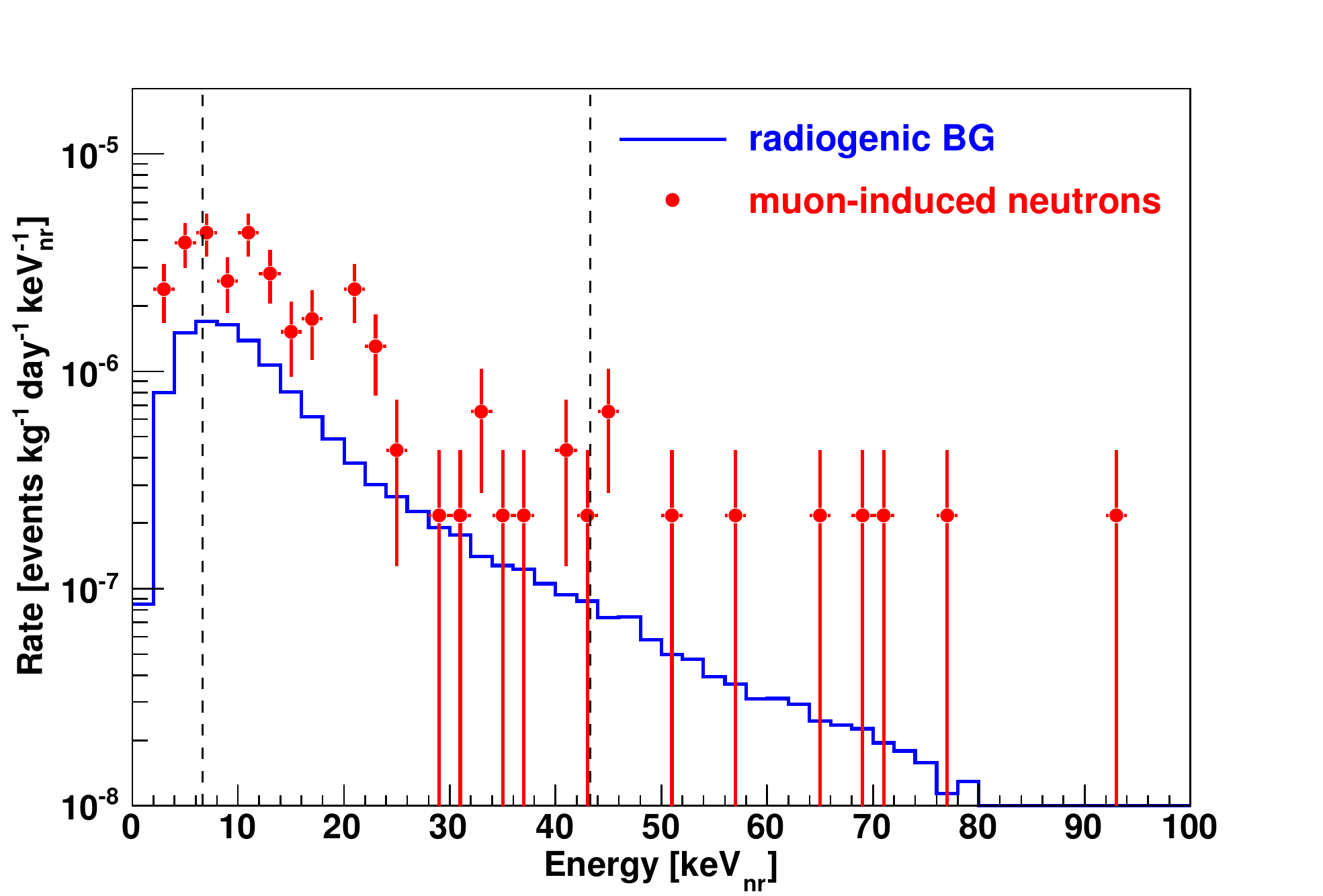}
\label{figTotalBG_1}}
\subfigure[~~converted to PE, with energy resolution and acceptance]{
\includegraphics[width=1.0\linewidth]{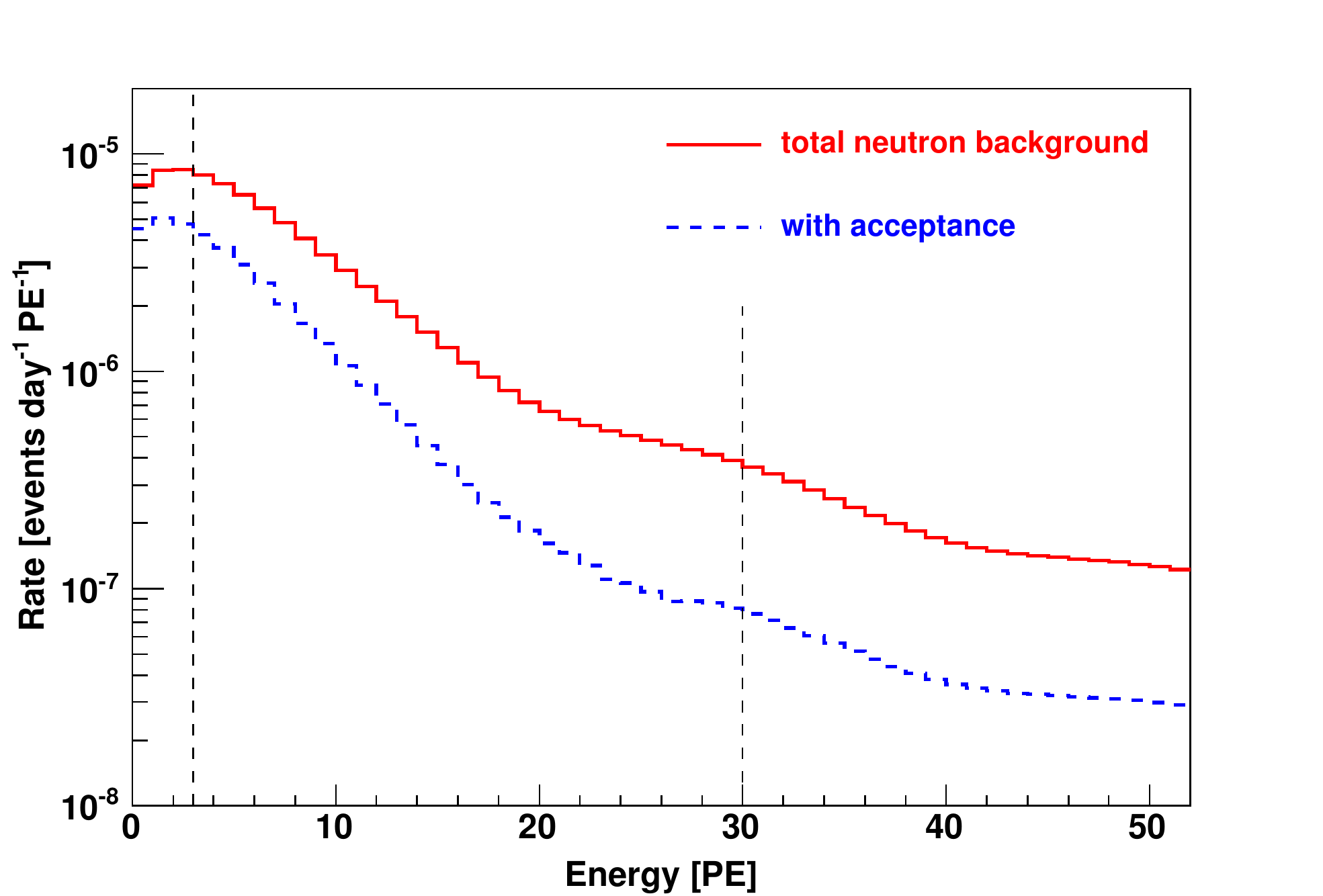}
\label{figTotalBG_2}}
\caption{Energy spectra of the NR background due to radiogenic and cosmogenic neutrons predicted for the 2012 WIMP search analysis (34~kg fiducial volume). The prediction takes into account the measured trigger threshold which is the cause of the roll-off at the lowest energies. The bottom plot shows the total NR background on the energy scale converted to PE, with the energy resolution and acceptance of the analysis cuts taken into account. Fluctuations of the light signal for events below the threshold have also been included in this analysis. The thin vertical dashed lines indicate the WIMP search energy region.}
\label{figTotalBG}
\end{figure}

The energy-dependent acceptance of these cuts was calculated with Monte Carlo simulations, taking into account the ${\cal L}_\mathrm{eff}$ parametrization~\cite{xe100-independent2}. The simulation results cannot be validated via direct comparison with the measured background data due to the very low rate of NRs. Hence, the results were verified by comparing the ratio of single and double scatter events using measured $^{241}$AmBe neutron calibration data and a corresponding Monte Carlo simulation, showing excellent agreement within the known uncertainties of ${\cal L}_\mathrm{eff}$ and the source strength~\cite{AmBeStudy}.

The total NR background rates predicted for the 2011~\cite{xe100-independent1} and the 2012~\cite{xe100-independent2} data analyses, taking into account the detection efficiency, the energy range, and the exposure, are given in Table~\ref{tabNRBGtot}. The contribution from environmental radioactivity (contamination in the rock and concrete of the LNGS laboratory) is at the percent level and has no significant impact on the XENON100 science goals. About 70\% comes from muon-induced neutrons. In future experiments, such as XENON1T~\cite{xenon1t}, this background will be significantly reduced by using a muon veto system. The energy spectra of the background from radiogenic and cosmogenic neutrons predicted for the results published in 2012 are shown in Fig.~\ref{figTotalBG_1}. Figure~\ref{figTotalBG_2} presents the total NR background expected in XENON100, converted to the S1 PE energy scale and corrected for the acceptance due to the analysis cuts, taking into account fluctuations of the light signal for events below the threshold.

\begin{table}[!tbp]
\centering
\caption{Nuclear recoil backgrounds predicted for the 2011~\cite{xe100-independent2} and the 2012~\cite{xe100-run10} WIMP searches, taking into account the detection efficiency. The NR acceptance after applying an ER discrimination cut based on S2/S1 ratio is not applied as it is not used in the standard Profile Likelihood analysis~\cite{ProfileLikelihood}.}
\label{tabNRBGtot}
\begin{tabular}{>\footnotesize{l} | >\footnotesize{c} | >\footnotesize{c}}
\hline
\hline

Data  release 		   					& 2011~\cite{xe100-independent2} 					& 2012~\cite{xe100-run10} 	\\
Live time 		   						& 100.9~days 									& 224.6~days 	\\
Fiducial volume 		     				& 48~kg 										& 34~kg 		\\
Energy range 		     					& 8.4$\--$44.6~keV$_{\mathrm{nr}}$				& 6.6$\--$43.3~keV$_{\mathrm{nr}}$ 		\\
									& (4$\--$30~PE)								& (3$\--$30~PE) 		\\

\hline
\multirow{2}{*}{Radiogenic neutrons}		& \multirow{2}{*}{0.10$\pm$0.02~events}	  	 		& \multirow{2}{*}{0.14$\pm$0.02~events}		\\
& & \\
Cosmogenic neutrons					& 0.21$^{+0.21}_{-0.12}$~events					& 0.34$^{+0.34}_{-0.17}$~events	\\
\hline
\multirow{2}{*}{Total NR background}		& \multirow{2}{*}{0.31$^{+0.22}_{-0.11}$~events}			& \multirow{2}{*}{0.48$^{+0.34}_{-0.17}$~events}	\\
& & \\
\hline
\hline
\end{tabular}
\end{table}

\section{Conclusions}
\label{secConclusions}
The NR background in the XENON100 experiment, originating from radiogenic and cosmogenic neutrons has been predicted for the XENON100 experiment based on Monte Carlo studies, using a detailed model of the detector and its shield. 

The total NR background in the 100.9~days dataset (2011, \cite{xe100-independent2}), which used a fiducial mass of 48~kg and (8.4$\--$44.6)~keV$_{\mathrm{nr}}$ energy range, is (0.31$^{+0.22}_{-0.11}$)~events. The detector's energy resolution as well as the active veto efficiency are taken into account here. The prediction for the 224.6~days data used for the WIMP search results of 2012~\cite{xe100-run10} is (0.48$^{+0.34}_{-0.17}$)~events. A fiducial target mass of 34~kg  and an energy range of  (6.6$\--$43.3)~keV$_{\mathrm{nr}}$ were used for this search.

With the reduced NR acceptance after applying the S2/S1 electronic recoil discrimination cut to define a benchmark WIMP search region, these values translate into (0.11$^{+0.08}_{-0.04}$)~events, and (0.17$^{+0.12}_{-0.07}$)~events, respectively. 
Compared to the total ER background estimates of (1.7$\pm$0.6)~events in the 2011 WIMP search, and (0.8$\pm$0.2)~events~\cite{xe100-independent2} in the result published in 2012, the neutron background does not limit the WIMP search sensitivity of the XENON100 experiment. 

The neutron background is even lower at energies above the upper bound used so far. In the (43.3$\--$100)~keV$_\mathrm{nr}$ energy range and 48~kg fiducial target, the contributions from radiogenic and muon-induced neutrons add up to (0.12$^{+0.12}_{-0.05}$) background events/year in the full discrimination space. In a 34~kg target it is only (0.07$^{+0.05}_{-0.04}$)~events/year, which is about 10\% of the background in the energy range below 30~PE.

~

\section{Acknowledgements}
We gratefully acknowledge support from NSF, DOE, SNF, UZH, FCT, R\'egion des Pays de la Loire, STCSM,  NSFC, DFG, Stichting voor Fundamenteel Onderzoek der Materie (FOM), the Max Planck Society and the Weizmann Institute of Science.

We are grateful to LNGS for hosting and supporting XENON100.

~

~

~

~

~

~

~

~

~

~

~

~

~

~

~

~

~

~
~

\end {document}